\documentclass[10pt,fleqn,a4paper]{article}
\pdfoutput=1
\usepackage{graphicx}
\usepackage{epsfig,cite}
\usepackage{amssymb}
\usepackage{amsmath}
\usepackage{color}
\usepackage{amstext,alltt,setspace}
\usepackage{amsbsy}
\usepackage{array}
\usepackage{booktabs,tabularx}
\usepackage{hyperref}
\usepackage{slashed}
\usepackage{url}
\usepackage{geometry}
\usepackage{
  pgf,
  tikz}
\usetikzlibrary{
  arrows,
  trees,
  scopes,
  decorations.text,
  decorations.pathreplacing,
  decorations.pathmorphing,
  decorations.markings,
  positioning,
  calc,
  patterns,
  intersections,
  shapes,
  fadings
}

\usetikzlibrary{trees}
\usetikzlibrary{decorations.pathmorphing}
\usetikzlibrary{decorations.markings}

\tikzset{
  photon/.style=
  {
    decorate,
    decoration={snake},
    draw=black
  },
  particle/.style=
  {
    very thick,
    draw=black,
    postaction={decorate},
    decoration=
    {
      markings,
      mark=at position .5 with {\arrow[draw=black]{>}}
    }
  },
  antiparticle/.style=
  {
    very thick,
    draw=black,
    postaction={decorate},
    decoration=
    {
      markings,
      mark=at position .5 with {\arrow[draw=black]{<}}
    }
  },
  gluon/.style=
  {
    decorate,
    draw=black,
    decoration=
    {coil,
      amplitude=4pt,
      segment length=5pt
    }
  },
  scalar/.style=
  {
    densely dashed,
    draw=black
  }
 }

\newcommand{\dd}{{\rm d}}

\newcommand{\order}[1]{{\cal O}(#1)}

\begin{document}

\begin{flushright}
  TIF-UNIMI-2019-3\\
\end{flushright}

\vspace*{.2cm}

\begin{center}
  {\Large \bf{Fitting the b-quark PDF as a massive-$b$ scheme:\\
      Higgs production in bottom fusion.}}
\end{center}

\vspace*{.7cm}

\begin{center}
  Stefano Forte$^{1}$, Tommaso Giani$^2$ and Davide  Napoletano$^{3}$
  \vspace*{.2cm}

{  \noindent
      {\it
        $^1$ Tif Lab, Dipartimento di Fisica, Universit\`a di Milano and\\ 
        INFN, Sezione di Milano,
        Via Celoria 16, I-20133 Milano, Italy\\
        $^2$ The Higgs Centre for Theoretical Physics, University of
        Edinburgh,\\ 
 JCMB, KB, Mayfield Rd, Edinburgh EH9 3JZ, Scotland\\
        $^3$ IPhT, CEA Saclay, CNRS UMR 3681,\\ F-91191, Gif-Sur-Yvette, France\\}}

      \vspace*{3cm}

      {\bf Abstract}
\end{center}

{\noindent
We show that a simple and accurate approach to the computation of
hadron collider processes involving initial-state $b$ quarks can be
obtained by  introducing an independently parametrized $b$ PDF. We
use the so-called FONLL method for the
matching of a scheme in which the $b$ quark is treated as a massless
parton to that in which it is treated as a massive state, and extend
it to the case
in which the $b$ quark PDF is not necessarily determined by
perturbative matching conditions. This generalizes to hadronic
collisions analogous results previously obtained for deep-inelastic
scattering. The results corresponds to a ``massive $b$''
scheme, in which $b$ mass effects are retained, yet the $b$ quark is
endowed with a PDF.
We 
specifically study Higgs production in bottom fusion, and show that
our approach overcomes difficulties related to the fact that in a
standard massive four-flavor scheme $b$-quark induced processes only
start at high perturbative orders.}

\pagebreak

\section{The treatment of heavy quark PDFs}
\label{sec:intro}

It has been recently shown~\cite{Ball:2017nwa} that for accurate
phenomenology at the LHC it is advantageous to treat the charm parton
distribution (PDF) on the same footing as light-quark PDFs, namely, to
parametrize it and extract it from data, rather than to take it
as radiatively generated from the gluon using  perturbative matching
conditions. This is likely to be
due to the fact that matching conditions are only known to
the lowest nontrivial order, which may well be subject to large higher
order
corrections, as revealed by the strong dependence of results on the
choice of matching scale. On top of this, of course,
the starting low-scale heavy quark
PDFs could in principle also have a
non-perturbative ``intrinsic''
component~\cite{Brodsky:1980pb,Brodsky:1981se}. It is important to note  that
whether or not the heavy quark PDF has a nonperturbative component,
and whether it is advantageous to parametrize the heavy quark PDF are
separate issues: in fact in Ref.~\cite{Ball:2017nwa} it was shown that
the main phenomenological
advantage in parametrizing and fitting the charm PDF comes from a region in
which any nonperturbative contribution to charm is likely to be
extremely small. 

The case of the bottom quark PDF is, in this respect, particularly
interesting. On the one hand, one may think that that the problem of
large higher order corrections to the matching conditions is alleviated
in this case by the larger value of the mass. However, on the other
hand, there is a more subtle consideration. Namely, there are $b$-initiated
hadron collider processes --- some of which
are especially relevant for new physics searches --- such as Higgs production in
bottom fusion, for which $b$ quark mass effects might be
non-negligible~\cite{Maltoni:2012pa,Lim:2016wjo,Bagnaschi:2018dnh}. This
suggests the use of a scheme in which the $b$ quark is
treated as a 
massive final-state parton --- hence not endowed with a
PDF. In such a scheme
the $b$-induced process necessarily
starts at a higher perturbative order than in  a scheme in which
there exists a $b$ PDF, because the $b$ production process is included in the
hard matrix element. As a consequence, the computation of the
$b$-induced process itself is more difficult and it can typically
only be performed with a lower perturbative accuracy than in a scheme
in which the $b$ quark is a massless parton.

The problem is somewhat
alleviated if the massive-scheme and massless-scheme
computations are combined, with the $b$-PDF in the massless scheme assumed to be
produced by perturbative matching conditions.  We  henceforth refer to
such a computational 
framework as ``matched-$b$''. However, in a matched-$b$ framework the
massive computation is still beset by the need to start at high
perturbative order.
As a possible way out, the use of a ``massive five flavor scheme'' has
been suggested recently~\cite{Krauss:2017wmx,Figueroa:2018chn}, in
which there is a $b$ PDF (hence five flavors), yet $b$ quark mass
effects are included (possibly, at least in part, also in parton showering).
The use of an independently parametrized $b$ quark PDF
within a framework in which massive and massless computations are
combined
offers a simpler way of dealing with the same
problem. We  refer to this as a ``parametrized $b$''
computational framework.
Such an approach has been developed for electroproduction in
Refs.~\cite{Ball:2015tna,Ball:2015dpa}, and it has been used in order
to produce PDF sets with parametrized
charm~\cite{Ball:2016neh,Ball:2017nwa}, including the recent NNPDF3.1
set. Because the only data currently used for PDF determination in which 
heavy quark mass effects have  a significant impact are deep-inelastic
scattering data close  to the charm production threshold, in these
references only electroproduction was considered and only the
parametrization of the charm was studied.

In these previous studies, an independently parametrized heavy quark PDF is
included in the so-called FONLL  method~\cite{Cacciari:1998it},
which allows for the
matching of a scheme in which the heavy quark mass is included but the
heavy quark decouples from QCD evolution equations, and a massless
scheme in which the heavy quark mass is neglected, but the heavy quark
PDF couples to perturbative evolution.
In this parametrized heavy quark version of the FONLL scheme, the heavy
quark PDF is present both in the massive and massless scheme, though
decoupled from evolution in the massive scheme; unlike in conventional
matched heavy quark computations
in which the number of PDFs is different, with one more
PDF in the massless scheme. The rationale for FONLL
with a parametrized heavy quark is to simultaneously include heavy quark
mass effects at lower scales and
the resummation of collinear mass logarithms in the heavy quark PDFs at
higher scales. This has the important byproduct that one
ends up with a computational framework in which there are heavy quarks in
the initial state even in the scheme in which mass effects are
retained.

Therefore, in a parametrized-$b$ FONLL framework, problems
related to the fact that the relevant processes in a massive scheme start
at higher order is thus completely
evaded, since the heavy quark 
PDF is always present. Mass effects are then included to finite
perturbative order, along with the resummation of mass logarithms, though
(unlike in some  ``massive five-flavor scheme'') mass corrections to
resummed perturbative evolution are not included. On the other hand,
any possible nonperturbative corrections to the $b$ PDF, including,
say, the effective value of the $b$ mass at which the matching should
happen, are then included in the PDF itself and thus extracted from
the data.

In this paper we explicitly construct the parametrized-$b$ FONLL method, by
generalizing to hadronic processes the construction 
of Refs.~\cite{Ball:2015tna,Ball:2015dpa} of FONLL  with
parametrized heavy quark PDF. We specifically consider the
application to Higgs production in bottom fusion. This process has
been computed at the matched level both using the FONLL
method~\cite{Forte:2015hba,Forte:2016sja} and EFT-based
methods~\cite{Bonvini:2015pxa,Bonvini:2016fgf}, with the respective results
benchmarked in
Ref.~\cite{deFlorian:2016spz} and found to be in good agreement with
each other. All these computations were performed
in a matched-$b$ approach, in which the $b$ PDF  is absent in the
massive (four-flavor) scheme, and determined by matching condition in
the massless (five-flavor) scheme. Here we  take this process as
a prototype for the use of a parametrized-$b$  scheme
for hadron-collider processes.

First, we discuss how the counting of perturbative orders changes
in the presence of a parametrized-$b$ PDF, and redefine suitable
matched schemes based on this new counting. We  then work out the
generalization to hadronic processes of FONLL with
parametrized heavy quark PDF of Refs.~\cite{Ball:2015tna,Ball:2015dpa},
we discuss in which sense it effectively provides an alternative to the
massive five-flavor scheme, and then we work out explicit expressions for
Higgs production in bottom fusion to the matched next-to-leading order
- next-to leading log (NLO-NLL) level and NLO-NNLL level. We 
finally compare
predictions obtained within this approach with some plausible choices
of the $b$-quark PDF to those obtained in the approach of
Refs.~\cite{Forte:2015hba,Forte:2016sja}, and argue that results with
similar or better phenomenological accuracy can be obtained in a much
simpler way within this new approach.

\section{The FONLL scheme with parametrized heavy quark PDF
  in hadronic collisions}
\label{sec:FONLL-HI}

Even though we have the general goal of constructing a parametrized-$b$
FONLL scheme for hadronic processes, we  always specifically refer to Higgs
production in gluon fusion, in order to have  a concrete reference
case, and  test scenario. We recall that the FONLL
method matches two calculations of the same process
performed in two different renormalization schemes: a massive scheme
in which the heavy quark mass is retained, but the heavy quark
decouples from the running of $\alpha_s$ and from QCD evolution
equations, and a massless scheme in which the heavy quark contributes
to the running of $\alpha_s$ and QCD evolution
equations, but the heavy quark mass is neglected.
In the computation of a hard process at scale $Q^2$, in the former scheme
mass effects $O\left(\frac {m_b^2}{Q^2}\right)$
are retained, but mass logarithms $\ln\frac{Q^2}{m_b^2}$
 are only
kept to finite order in $\alpha_s$ (where $m_b$ denotes generically
the mass of the heavy quark). In the latter scheme, mass effects
are neglected but mass logarithms are resummed to all orders in
$\alpha_s$. Hence by matching the two calculations one retains
accuracy both at low scales where quark mass effects are important, and
at high scales where mass logarithms are large.

The general idea of the FONLL method is to realize that these are just
two different renormalization schemes: the massive scheme is a
decoupling scheme, and the massless scheme is a minimal subtraction
scheme. So the two calculations can be simply matched by re-expressing
both in the same renormalization scheme, and then subtracting common
contributions. In practice, this is done by expressing the massive
scheme computation in terms of the PDFs and $\alpha_s$ of the
massless scheme, and then adding to it the difference $\sigma ^{\rm
  d}$
between the massless calculation and the massless limit of the
massive one. Schematically
\begin{align}\label{eq:fonlldef}
&  \sigma^{\text{FONLL}} = \sigma^{\rm massive} +\sigma ^{\rm d}\\
&\qquad \sigma ^{\rm d}=\sigma^{\rm massless}-\sigma^{\rm massive,\,0}.\label{eq:fonlldiff}
\end{align}
This corresponds to replacing all the terms in the massless
computation which are included to finite order in $\alpha_s$ in the
massive computation with their massive counterpart.

In the simplest (original) version of  FONLL, as discussed
in Ref.~\cite{Cacciari:1998it} for $b$ production in hadronic
collisions, and in Ref.~\cite{Forte:2010ta} for deep-inelastic
scattering, in the massive scheme there is no heavy quark PDF, and the
 heavy quark can only appear as a final-state particle. In the
 massless scheme the heavy quark PDF is determined by matching
 conditions in terms of the light quarks and the gluon. These
 conditions match the massless scheme at a scale $\mu$ such that the
 heavy quark PDF only appears for scales above $\mu$. Specifically,
 at order $O(\alpha_s)$, the heavy quark PDF just vanishes
 at the scale 
 $\mu=m_b$
 and it is
 generated by perturbative evolution at higher scales, while at
 $O(\alpha^2_s)$ it has a nontrivial gluon-induced matching condition
 at all scales.

 When introducing a parametrized PDF both the massive and massless
 scheme computations change. The massless scheme changes, somewhat
 trivially,
 in that the
 heavy quark PDF, at the matching scale, instead of being given by a
 matching condition,  is freely parametrized. The massive scheme
 changes nontrivially in that there is now a heavy quark PDF also in
 this scheme, 
only it does not evolve with the scale.  The
 consequences of this were worked out in
 Refs.~\cite{Ball:2015tna,Ball:2015dpa} in the case of
 electroproduction, and we study them here for hadroproduction for the
 first time.

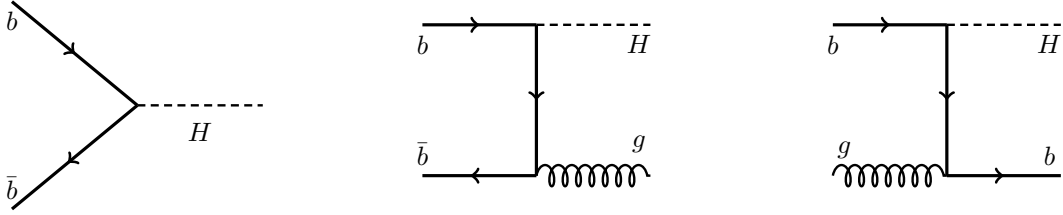
\begin{figure}[tbp]
  \centering{
    \begin{minipage}{0.32\linewidth}
      \begin{tikzpicture}[line width=0.9 pt, scale=1.1]
        \draw[particle] (-1.5,1.25) -- (0,0) node[pos=0.0,below]{$b$}; ;
        \draw[antiparticle] (-1.5,-1.25) -- (0,0) node[pos=0.0,above]{$\bar{b}$};
        \draw[scalar] (0,0) -- (1.5,0) node[midway,below=0.1cm] {$H$} ;
      \end{tikzpicture}
    \end{minipage}
    \begin{minipage}{0.32\linewidth}
      \begin{tikzpicture}[line width=0.9 pt, scale=1]
        \draw[particle] (-1.5,1.0) -- (0,1.) node[pos=0.,below] {$b$};
        \draw[antiparticle] (-1.5,-1.) -- (0,-1.) node[pos=0.0,above] {$\bar{b}$};
        \draw[particle] (0,1) -- (0.,-1);
        \draw[gluon] (0.,-1) -- (1.5,-1) node[pos=.9,above=5pt] {$g$};
        \draw[scalar] (0.,1) -- (1.5,1) node[pos=.9,below] {$H$};
      \end{tikzpicture}
    \end{minipage}
    \begin{minipage}{0.32\linewidth}
      \begin{tikzpicture}[line width=0.9 pt, scale=1]
        \draw[particle] (-1.5,1.0) -- (0,1.) node[pos=0.,below] {$b$};
        \draw[gluon] (-1.5,-1.) -- (0,-1.) node[pos=0.1,above=3pt] {$g$};
        \draw[particle] (0,1) -- (0.,-1);
        \draw[particle] (0.,-1) -- (1.5,-1) node[pos=.9,above] {$b$};
        \draw[scalar] (0.,1) -- (1.5,1) node[pos=.9,below] {$H$};
      \end{tikzpicture}
    \end{minipage}
  }
  \caption{Feynman diagrams for the leading (left) and next-to-leading order real emission
    contributions to  Higgs production in bottom fusion which are present
    in the massive scheme when the $b$ quark PDF is independently
    parametrized, but absent otherwise.}
  \label{fig:massive-4fs}
\end{figure}

\subsection{Perturbative ordering}
\label{sec:pertord}

Because there is now a $b$ PDF also in the massive scheme, the
counting of perturbative orders in this scheme
changes substantially. Specifically, for Higgs production in bottom
fusion the diagrams of Fig.~\ref{fig:massive-4fs} are present only
when the $b$ PDF is independently parametrized. This means that while 
in the
massive scheme the process in the matched-$b$ approach of
Refs.~\cite{Forte:2015hba,Forte:2016sja}  starts at $O(\alpha_s^2)$,
in a parametrized-$b$ approach it starts at 
$O(\alpha_s^0)$. As discussed in detail in
Refs.~\cite{Forte:2010ta,Forte:2015hba,Forte:2016sja},
the FONLL method allows the  consistent combination of computations performed
at different perturbative orders either in the massive or massless
scheme: various combinations were defined and discussed  in
Refs.~\cite{Forte:2015hba,Forte:2016sja} for Higgs production in
bottom fusion.

With the new counting of perturbative orders which is relevant for a
parametrized-$b$ framework it is convenient to define some new
combinations. We  consider in particular the combination of 
the massive scheme $O(\alpha_s)$ computation  with the standard
five-flavor scheme next-to-leading log (NLL) and
next-to-next-to-leading log  computations. We  call these
combinations respectively FONLL-AP (hence corresponding to NLO-NLL)
and FONLL-BP (corresponding to  NLO-NNLL).

\subsection{Parametrized-$b$ FONLL}
\label{sec:parbfonll}

The construction of the parametrized-$b$ FONLL for hadronic processes
closely follows the corresponding construction for electroproduction,
presented in Refs.~\cite{Ball:2015tna,Ball:2015dpa}, to which we refer
for more details. In comparison to the matched-$b$ FONLL of
Refs.~\cite{Forte:2015hba,Forte:2016sja} the massive scheme
contribution to Eq.~(\ref{eq:fonlldef}) includes an extra contribution:
\begin{align}
  \label{eq:delta-i}
  \sigma_{\text{FONLL}_P} &= \sigma_{\text{FONLL}_M} +
  \delta\sigma_P\nonumber\\ 
  \quad \quad \delta\sigma_P &= \sigma^{\rm massive}_{P} - \sigma^{{\rm
      massive},0}_{P},
\end{align}
where $\sigma^{\rm massive}_{P}$ is the massive-scheme contribution to
the given process with initial-state heavy quarks and $\sigma^{{\rm
      massive},0}_{P}$ its massless limit (which subtract its double
counting with the massless-scheme contribution). This massive scheme
contribution has to be re-expressed in terms of massless-scheme PDFs,
as explained in detail in
Refs.~\cite{Cacciari:1998it,Forte:2010ta,Ball:2015tna,Ball:2015dpa,Forte:2015hba,Forte:2016sja}.

For Higgs production in bottom fusion, up to NLO, this extra
contribution is given by
the real diagrams of Fig.~\ref{fig:massive-4fs}, supplemented
by the corresponding virtual correction and thus it
has the form
\begin{equation}
  \label{eq:xs4fs-i}
  \begin{split}
    \delta\sigma_P^{\rm massive}\left(\frac{m_H^2}{m_b^2}\right) = &
    2\,\int_{\tau_0}^1 \frac{\dd{x}}{x}\,
    \int_{\frac{\tau_0}{x}}^1 \frac{\dd{y}}{y^2}\,
    f^{(4)}_b(x)\,f^{(4)}_{\bar{b}}\left(\frac{\tau_0}{xy}\right)
    \left[\sigma_{b\bar{b}}^{(4),(0)}\left(y,\frac{m_H^2}{m_b^2}\right) +
      \alpha_s\,\sigma_{b\bar{b}}^{(4),(1)}\left(y,\frac{m_H^2}{m_b^2}\right)
    \right]\\
    &
    +
    4\,\alpha_s\,f^{(4)}_b(x)\,f^{(4)}_g\left(\frac{\tau_0}{xy}\right)\,
    \sigma_{bg}^{(4),(1)}\left(y,\frac{m_H^2}{m_b^2}\right)\,,
  \end{split}
\end{equation}
where the subscript $P$ denotes the fact that this contribution is
only present when the $b$ PDF is independently parametrized, and the
superscript $(4)$ is used to denote the massive factorization scheme,
as in Refs.~\cite{Forte:2015hba,Forte:2016sja}. Note that even though,
with a parametrized $b$ there are five flavors also in
the massive scheme, only the four lightest ones contribute to
the running of $\alpha_s$ and perturbative evolution. The
massive cross-sections $\sigma_{ij}^{(4),(k)}$ were computed e.g. in
Ref.~\cite{Krauss:2017wmx} based on corresponding QED results from
Ref.~\cite{Dittmaier:1999mb} and are collected in
Appendix~\ref{sec:app-coeff} after scheme change as we discuss below.

Note that in the matched-$b$ computation
of Ref.~\cite{Forte:2015hba,Forte:2016sja} this
process  in the massive scheme starts at $O(\alpha_s^2)$, hence up to NLO (with our new
counting) the contribution given in Eq.~(\ref{eq:xs4fs-i}) is the only
one to $\sigma^{\rm massive}$ Eq.~(\ref{eq:fonlldef}): so in actual fact
in this case
\begin{equation}
   \sigma^{\rm massive,\,NLO}= \sigma^{\rm massive,\,NLO}_{P}.
\end{equation}

The expression of $\sigma^{\rm massive,\,NLO}$ suitable for use in the
FONLL formula Eq.~(\ref{eq:fonlldef}) is obtained, as mentioned, by
  re-expressing the massive scheme PDFs  and $\alpha_s$ in terms of massless-scheme
  ones. For simplicity we assume that this is done at a matching scale
  $\mu_b=m_b$.
  The matching condition for $\alpha_s$ is, as well known,
\begin{equation} \label{eq:amatch}
  \alpha_s^{(4)}(Q^2) = \alpha^{(5)}_s(Q^2)\left[1\,-\alpha_s\,\frac{T_R}{2\,\pi}\,\log\frac{Q^2}{m_b^2} + \order{\alpha_s^2}\right]
\end{equation}
while to $O(\alpha_s)$ the only nontrivial matching condition is
  that for the $b$ PDF:
\begin{equation}
  \label{eq:b-pdf}
  f_b^{(4)}(x) = f^{(5)}_b(x,Q^2) - \alpha_s\int_x^1 \frac{\dd{z}}{z}
  \left[ K_{bb}^{(1)}\left(z,Q^2\right)f^{(5)}_b\left(\frac{x}{z},Q^2\right)
   + K_{bg}^{(1)}(z,Q^2)\,f_g^{(5)}\left(\frac{x}{z},Q^2\right)\right]+ \order{\alpha_s^2},
\end{equation}
where again the superscripts $(4)$ and  $(5)$  denote  respectively the massive-
and massless-scheme expressions, and $K_{ij}$ are the matching
coefficients
\begin{equation}\label{eq:matching}
  f_i^{(5)}(Q^2) = \sum_j K_{ij}(Q^2) \otimes f^{(4)}_j(Q^2),
\end{equation}
where the sum runs over all partons (including the heavy quark),
and
\begin{equation}\label{eq:kexp}
  K_{ij}(Q^2)=\delta_{ij}\delta(1-z)+\sum_{n=1}\alpha^n_s(Q^2) K_{ij}^{(n)}(Q^2).
\end{equation}
Note
that, of course, since there is a heavy quark PDF also in the massive
scheme, $K_{ij}$ is a square matrix, so that, to $O(\alpha_s)$,
$  K^{-1}_{ij}(Q^2)=\delta_{ij}-\alpha_s(Q^2) K_{ij}^{(1)}(Q^2)$.
The matching function 
 $K^{(1)}_{bb}$ was calculated in Ref.~\cite{Buza:1996wv}. Its explicit
expression is given for ease of reference in Appendix~\ref{sec:app-splitting} together with
that of the splitting functions
$P_{ij}$. 

Substituting Eq.s~(\ref{eq:amatch}-\ref{eq:b-pdf}) in
Eq.~(\ref{eq:xs4fs-i}) we get the desired expression for the
massive-scheme cross section:
\begin{equation}
  \label{massive:1} \sigma_P^{\rm
    massive}\left(\frac{m_H^2}{m_b^2}\right)=
  \int_{\tau_H}^{1} \frac{dx}{x}\int_{\frac{\tau_H}{x}}^{1}
  \frac{dy}{y^2}\sum_{ij=b,g}f_{i}^{(5)}(x,Q^2)f_j^{(5)}
  \left(\frac{\tau_H}{x y},Q^2\right)
  B_{ij}\left(y,\alpha_s^{(5)}(Q^2),
    \frac{Q^2}{m_b^2}\right),
\end{equation}
where to $O(\alpha_s)$ the non-vanishing coefficients are 
\begin{align}
  \label{eq:change-of-scheme}
  B_{b\bar{b}}^{(0)}\left(y,\frac{m_H^2}{m_b^2}\right)  
    & = \sigma_{b\bar{b}}^{(4),(0)}\left(y,\frac{m_H^2}{m_b^2}\right) \\
  B_{b\bar{b}}^{(1)}\left(y,\frac{m_H^2}{m_b^2}\right)
    & = \sigma_{b\bar{b}}^{(4),(1)}\left(y,\frac{m_H^2}{m_b^2}\right)-
      2\,\sigma_0
      \int_{y}^{1}{\rm d} z \,z\,\delta(z-y)\,
      K_{bb}^{(1)}\left(z,\ln\frac{m_H^2}{m_b^2}\right)\\
  B_{bg}^{(1)}\left(y,\frac{m_H^2}{m_b^2}\right)
    & = \sigma_{bg}^{(4),(1)}\left(y,\frac{m_H^2}{m_b^2}\right)-
      \sigma_0
      \int_{y}^{1}{\rm d} z \,z\,\delta(z-y)\,
      K_{bg}^{(1)}\left(z,\ln\frac{m_H^2}{m_b^2}\right)\,,
\end{align}
  whose explicit expressions are collected, as mentioned, in
  Appendix~\ref{sec:app-coeff}.

In order to construct the FONLL expression Eq.~(\ref{eq:fonlldef}), the
massive scheme contribution must be combined with the difference term
$\sigma ^{\rm d}$ Eq.~(\ref{eq:fonlldiff}). However, it is easy to
check that, just like in the case of
electroproduction\cite{Ball:2015tna,Ball:2015dpa},
this term, which is subleading when using matched $b$, vanishes
identically with parametrized $b$. This is due to the
fact that the massless limit of the massive-scheme calculation only
differs from the massless-scheme calculation because of the resummation
of mass logarithms $\ln\frac{Q^2}{m_b^2}$ beyond the accuracy of the
massive-scheme result (so at $O(\alpha_s^2)$ and beyond, in our
case). However, when re-expressing the massive-scheme calculation in
terms of massless-scheme PDFs the evolution of the $b$-PDF is only
removed up to the same accuracy as that of the massive scheme
calculation. This is seen explicitly in Eq.~(\ref{eq:b-pdf}), in which
mass logarithms  $\ln\frac{Q^2}{m_b^2}$ are only removed up to
$O(\alpha_s)$. Therefore, the higher order logarithms remain unsubtracted in
the expression of $f_b^{(5)} (x,Q^2)$ and thus cancel exactly between
$\sigma^{\rm massless}$ and $\sigma^{\rm massive,\,0}$.

The FONLL
result thus reduces to the expression Eq.~(\ref{massive:1}):
\begin{equation}\label{eq:canc}
\sigma^{\text{FONLL-AP}}=\sigma_P^{\rm
  massive}\left(\frac{m_H^2}{m_b^2}\right).
\end{equation}
We can thus write the FONLL result in the form
\begin{equation}\label{eq:genfonll}
\sigma^{\text{FONLL-AP}}= \sum_{i,j}\sum_{l,m} \sigma^{\rm massive}_{ij}\left(\frac{m_h^2}{m^2_b}\right)\otimes
 K^{-1}_{il}\otimes f_{l}^{\left(5\right)}\left(Q^2\right)K^{-1}_{jm}\otimes f_{m}^{\left(5\right)}\left(Q^2\right),
\end{equation}
where $ K^{-1}_{il}$ is the inverse of the matching matrix defined in
Eq.~(\ref{eq:amatch}), perturbatively defined order by order according
to Eq.~(\ref{eq:kexp}). This is
of course  well defined with a parametrized $b$ because
$K_{ij}$ is  a square matrix. As discussed in detail in
Refs.~\cite{Ball:2015tna,Ball:2015dpa} the effect of the inverse
matching matrices in Eq.~(\ref{eq:genfonll}) is to remove collinear
logarithms related to the evolution of the $b$ PDF from the massless scheme
PDFs $f_{i}^{\left(5\right)}$, since these are already included in the
massive-scheme matrix cross-section  $\sigma^{\rm massive}_{ij}$,
where they would appear as mass logarithms $\ln\frac{Q^2}{m_b^2}$
in the large $Q^2$ limit (in actual fact here $Q^2=m_H^2$). As a
consequence, the result Eq.~(\ref{eq:genfonll}) is completely
independent of the matching scale $m_b$ (i.e. the scale at which the
$b$ PDF is parametrized), as it must be, given that once
the $b$ PDF is parametrized there is no matching scale left.
We will check this cancellation explicitly (see 
Fig.~\ref{fig:mub-var} below).

Equation~(\ref{eq:genfonll}) shows that FONLL effectively acts
as a massive five-flavor scheme, in which standard five-flavor PDFs
are combined with the massive-scheme cross-section, with massive
quarks included in the initial state: it is in fact akin to 
five-flavor scheme of Ref.~\cite{Krauss:2017wmx}, though in this
reference mass effects are also included in parton showering, which we
do not consider here. In FONLL as corrections are consistently
included to the order at which the massive-scheme cross-section is
computed, with collinear and mass logarithms resummed to the logarithmic
order to which PDFs are used. The structure of the result
Eq.~(\ref{eq:genfonll}) is universal, and so are the PDFs which appear
in it. Therefore, to the extent that the  PDF is correctly fitted,
mass corrections (i.e. all terms suppressed by powers of $m_b/Q$) are
then fully included  up to the order of the massive-scheme
calculation:   $O(\alpha_s)$ for FONLL-AP and FONLL-BP. Of course
these latter corrections are not universal and will have to be computed
separately for each process.

As mentioned, the FONLL framework
allows for the combination of massive- and massless-scheme
computations performed at arbitrary, independent accuracy. We discuss
specifically the two cases defined in Sect.~\ref{sec:pertord}, FONLL-AP and
FONLL-BP. In FONLL-AP, the massive-scheme partonic cross sections 
 $\sigma_{ij}^{\rm massive}$ are computed up to NLO
(i.e. $O(\alpha_s)$), while the PDFs are evolved using 
NLO (more properly, NLL) evolution equations. Hence, in this case
Eq.~(\ref{eq:genfonll}), with  $\sigma_{ij}^{\rm massive}$ computed up
to $O(\alpha_s)$ (i.e. including the diagrams of
Fig.~\ref{fig:massive-4fs}), and NLO PDFs.

In FONLL-BP, the
massive-scheme computation is still performed up to NLO, but now the
massless-scheme computation is performed up to NNLO. This has two
consequences. The first is that NNLO PDFs are now used. The other is
that hard cross-sections are now computed up to NNLO i.e. up to
$O(\alpha^2_s)$. Because massive terms are included only up to
$O(\alpha_s)$, Eq.~(\ref{eq:genfonll}) must now be supplemented by a
purely massless $O(\alpha^2_s)$ contribution:
\begin{equation}\label{eq:fonll-bp}
\sigma^{\text{FONLL-BP}}=\sigma^{\text{FONLL-AP}}+\sum_{l,m}
\sigma^{(5),(2)}_{lm}\otimes
f_{l}^{\left(5\right)}\left(Q^2\right)f_{m}^{\left(5\right)}\left(Q^2\right), 
\end{equation}
where $\sigma^{\text{FONLL-AP}}$ is given by
Eq.~(\ref{eq:genfonll}). Note that because the matching functions
$K_{ij}^{-1}$ are 
used to re-express the massive-flavor scheme cross-section in the
massless scheme, they are accordingly  computed to the same accuracy as
the massive-scheme partonic cross-section itself: so here
to $O(\alpha_s)$. The difference term $\sigma^{\rm d}$
Eq.~(\ref{eq:fonlldiff}) always vanishes identically. It is clear
that the computation is considerably streamlined in comparison to the
standard FONLL framework of Refs.~\cite{Forte:2015hba,Forte:2016sja}.

\section{Higgs production in $b$ fusion}
\label{sec:results}

We now present explicit results for Higgs production in $b$-quark fusion
within the FONLL-AP and FONLL-BP scheme, and compare them to previous
results of Refs.~\cite{Forte:2015hba,Forte:2016sja}.
Results presented in this section are obtained using
the following set-up.
For the calculation of the 5F scheme coefficient functions,
we use the interface to the {\verb bbh@nnlo }
code~\cite{Harlander:2002wh}
as implemented in the public {\tt bbhfonll} code~\cite{code}.
The subtraction terms needed for the FONLL-B calculation of
Refs.~\cite{Forte:2015hba,Forte:2016sja} 
is obtained using  {\tt bbhfonll}. The standard contributions
to the 4F scheme are computed
using the MG5\_aMC@NLO package~\cite{Alwall:2014hca,Wiesemann:2014ioa},
while we have implemented the new terms $\delta\sigma_P^{\rm massive}$
Eq.~(\ref{eq:xs4fs-i})
and their
massless limit in a new version of  {\tt bbhfonll}, following
the expressions reported in Appendix~\ref{sec:app-coeff}, and is
available at
\begin{center}
\url{https://gitlab.com/dnapoletano/bbh-intrinsic-public}.
\end{center}
Both codes use the LHAPDF~\cite{Buckley:2014ana} package.

We use the  NNPDF3.1 NNLO set of parton distributions with
$\alpha_s(M_z)=0.118$~\cite{Ball:2017nwa}. For a first default
comparison we just use the vanilla NNPDF3.1 set, including the $b$
PDF. From the point of view of a computational framework in which the
$b$ PDF is fitted,
this can be thought of as the $b$ PDF that one would get if initial
PDFs were parametrized at $Q_0=m_b$, and the fitted $b$ PDF were to
turn out to be exactly equal to that given by the matching condition
at this scale. Furthermore, in order
to get a feeling for effects related to the size of
the $b$-PDF we then consider, for the sake of argument, a $b$ PDF
equal to that which would be obtained by using the matching condition
at $\mu_b = 2/3 m_b$ or $\mu_b = 1/2 m_b$, and then evolving up to
$Q=m_b$ where the initial PDF is given.


\begin{figure}[htbp]
  \centering
  \includegraphics[width=0.6\textwidth]{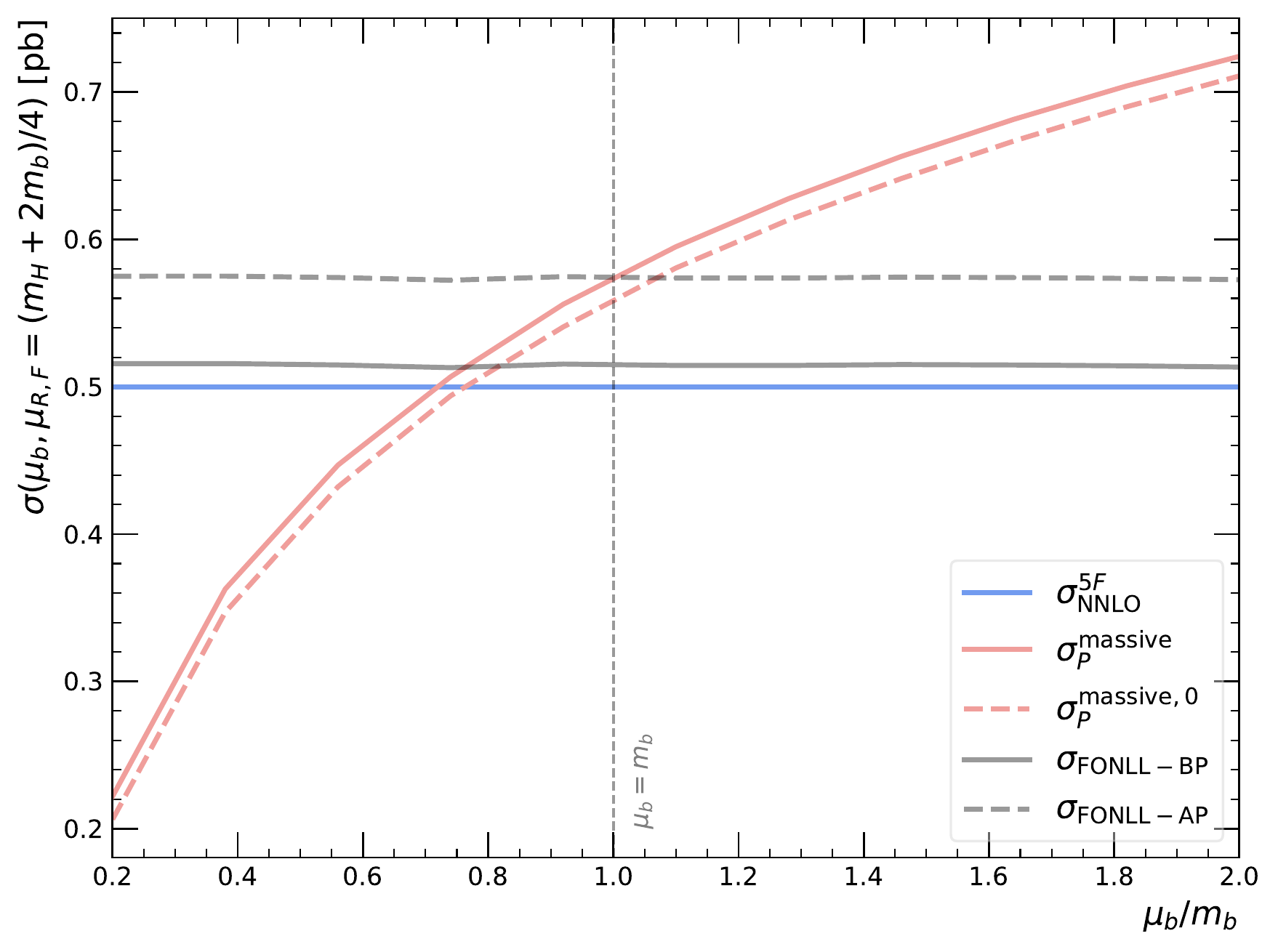}
  \caption{Cancellation of the dependence on the matching scale in the
  FONLL-AP and FONLL-BP schemes.}
  \label{fig:mub-var}
\end{figure}
First, as a consistency check, in Fig.~\ref{fig:mub-var} we verify that
indeed the dependence on $\mu_b$ cancels when constructing the FONLL
result with parametrized $b$ according to Eq.~(\ref{eq:delta-i}). In
this figure the massive-scheme result has been constructed using a
fixed  $b$ PDF (that which corresponds to the standard matching
condition at $\mu_b=m_b$) and then re-expressing results in terms of
the massive scheme PDFs  and $\alpha_s$ in terms of massless-scheme
ones. This is done using Eq.~(\ref{eq:matching}), which contains the
matching coefficients $K_{ij}$ which depend on the matching scale
$\mu_b$, and thus the massive-scheme result becomes
$\mu_b$-dependent.
However, this dependence cancels exactly in the final
FONLL result.

\begin{figure}[htbp]
  \centering
  \includegraphics[width=0.49\linewidth]{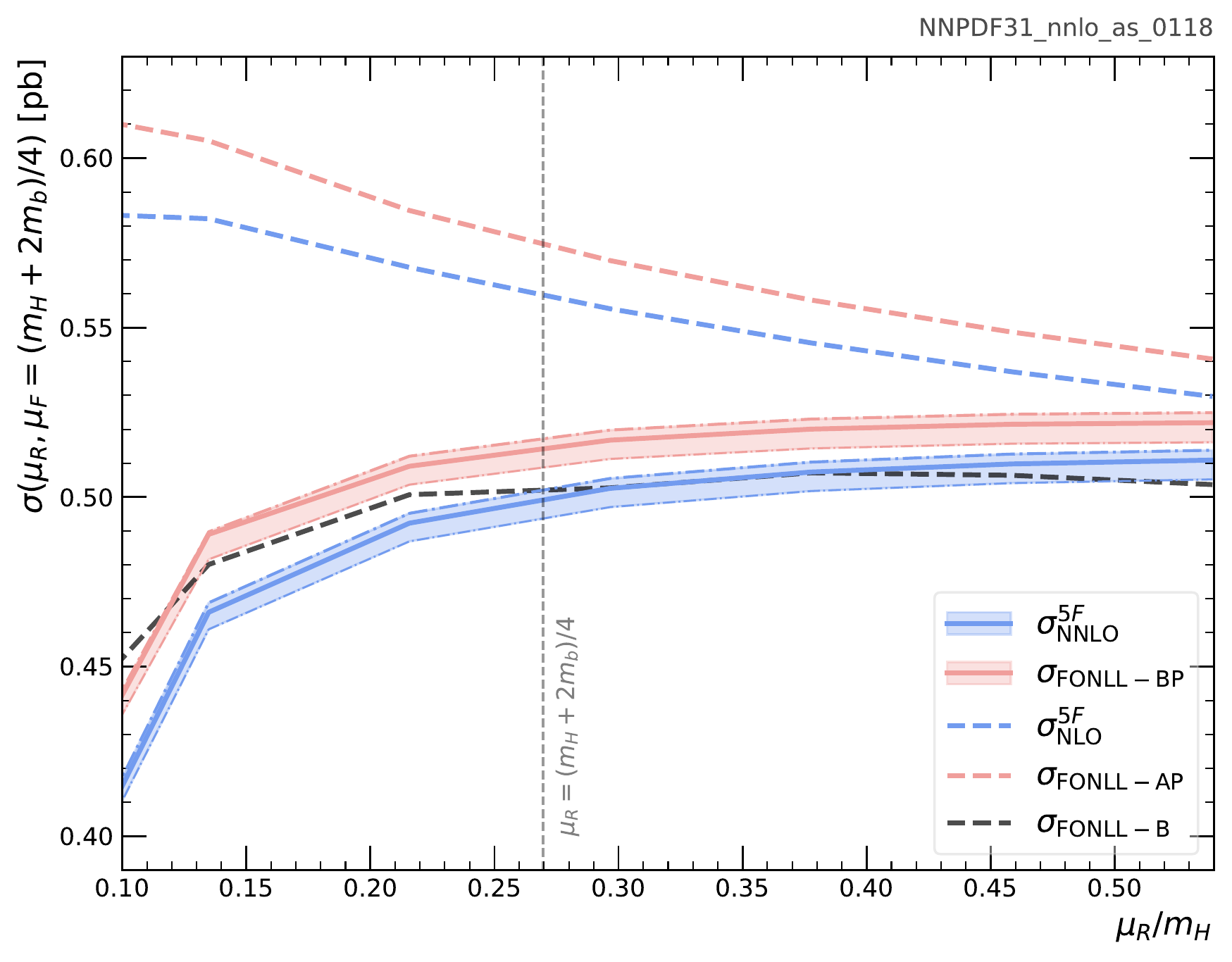}
  \includegraphics[width=0.49\linewidth]{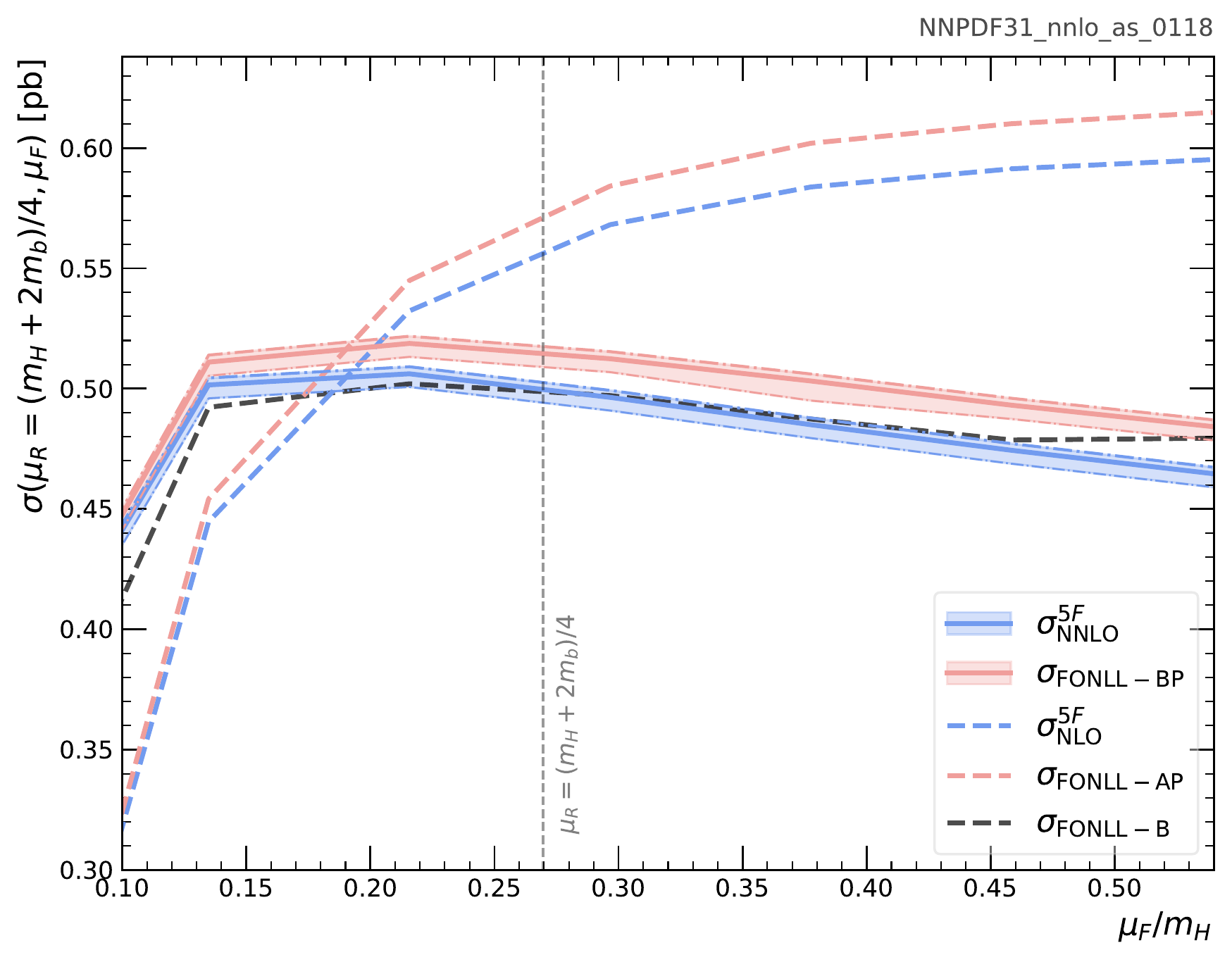}
  \caption{Renormalization (left) and factorization (right) scale
    variation of the cross-section for Higgs production in bottom
    fusion. The pure five-flavor scheme computation is compared to the
  FONLL-AP and FONLL-BP results presented here and to the FONLL-B
  result of Ref.~\cite{Forte:2015hba}. For the pure five flavor
  NNLO and the  FONLL-BP three curves are shown, corresponding to 
  three choices of initial $b$ PDF (see text).
  }
  \label{fig:scale-var}
\end{figure}

In Fig.~\ref{fig:scale-var} we show the factorization and
renormalization scale dependence of the cross-section computed in
various schemes, with the other scale kept fixed at the
preferred~\cite{Forte:2015hba,Forte:2016sja} value
$\mu=\frac{m_H+2m_b}{4}$.
Specifically, we compare results obtained using the FONLL-AP and
FONLL-BP schemes discussed above, the pure five-flavor scheme , and
the FONLL-B result of Ref.~\cite{Forte:2016sja}, all using the same
PDFs (including the $b$ PDF) as discussed above. For the pure
five-flavor scheme and for FONLL-BP we also show
the three curves corresponding to the three different choices for the
$b$ PDF discussed above, with a corresponding band:
the central, thick, solid line represents the default $\mu_b=m_b$
choice, while the edges of the band are drawn with dot-dash curves
with decreasing thickness, with the thicker of the two corresponding
to $\mu_b=2\frac{m_b}{3}$, and the other two $\mu_b=\frac{m_b}{2}$.

Note that the
FONLL-BP computation Eq.~(\ref{eq:fonll-bp}) and the FONLL-B
result~\cite{Forte:2016sja} are directly comparable: indeed, they both
include the five-flavor scheme computation up to NNLO, and combine it
with the first
two orders of the massive-scheme computation. The difference is that
in FONLL-B in the massive-scheme computation refers to the process
$gg\to b\bar b H$, while in FONLL-BP it refers to  $b\bar b \to H$. If
the $b$ PDF is the same as given by perturbative matching, the
difference is then only that, in the latter case, only mass effects
related to the $b\bar b$ which fuses into the Higgs are included,
while in the former, also those related to the further unobserved
$b\bar b$ pair are present. In a realistic situation, in which FONLL-BP is used
while parametrizing and fitting the $b$ PDF these mass effects should
be reabsorbed in the fitted $b$ PDF. In our comparisons, they 
appear as a certain enhancement of FONLL-BP in comparison to FONLL-B
due to the opening of phase space.

Otherwise, the qualitative features of the comparison between FONLL
and the pure five-flavor scheme remain essentially the same as discussed
in Ref.~\cite{Forte:2016sja}: FONLL is quite close to the
five-flavor scheme, with mass effects a non-negligible, but small,
positive correction. Indeed, the difference between
FONLL-AP and FONLL-BP, i.e., the impact of NNLO corrections in the
five-flavor scheme, is much more significant than that of mass corrections.
The impact of varying the $b$ PDF by
an amount which is comparable to a reasonable variation of the
matching scale is clearly comparable to that of the mass
corrections. This provides evidence for the fact that fitting the $b$
PDF is likely to have a significant impact on precision phenomenology.

Note that results for the FONLL-B
scheme differ at the percent level from those of 
Ref.~\cite{Forte:2016sja} because there a different PDF set and $m_b$
value were used,
for the sake of benchmarking with
Ref.~\cite{Bonvini:2015pxa,Bonvini:2016fgf}. This further highlights
the fact that the size of effects due to the $b$ PDF is comparable to
that of mass corrections.

\section{Conclusions}
\label{sec:conclusions}

In summary, we have shown how the FONLL matching of massive- and
massless-scheme treatment of computations involving heavy quarks can
be generalized to the case in which the heavy quark PDF is freely
parametrized for hadronic processes. We have show that this
effectively provides us with a massive heavy quark scheme, in which
the heavy quark is endowed with a standard PDF satisfying QCD
evolution equations, yet it is treated as massive in hard matrix
elements. A first application to Higgs production in bottom fusion
shows that effects related to the $b$ PDF are quite likely to be
comparable to mass corrections: both are small, but non-negligible
corrections to a purely massless NNLO calculation in which the $b$ PDF
is obtained from perturbative matching conditions. Determining the $b$
PDF from data is thus likely to be necessary for a description of
$b$-induced hadron collider processes at percent or sub-percent accuracy.

As a direction for further study, it should be noticed that extending
our results to NNLO --- thereby allowing the construction of a
FONLL-CP result, in the terminology of Sect.~\ref{sec:pertord}
(NNLO+NNLL) --- is beyond current knowledge. Indeed, starting at NNLO
the cancellation between real and virtual
corrections is no longer trivial, and is spoiled by
massive quarks in the initial
state~\cite{Doria:1980ak,Catani:1985xt}. Hence, such an extension
would require conceptual advances in the understanding of QCD
factorization in the presence of massive quarks, which are left for
future studies.

\section*{Acknowledgments}
We thank  Maria Ubiali for technical help and
stimulating discussions, and 
Valerio Bertone for support with the APFEL code. Also, we would
like to thank Marco Zaro for his assistance in using MG5\_aMC@NLO.
DN would like to thank the CERN theory department for hospitality
during the latter stages of this work.
The work of DN is supported by the French Agence Nationale de
la Recherche, under grant ANR-15-CE31-0016.
SF is supported by the European Research Council under
the European Union's Horizon 2020 research and innovation Programme
(grant agreement n.740006). TG is supported by The Scottish Funding
Council, grant H14027.

\appendix
\numberwithin{equation}{section}
\setcounter{equation}{0}

\section{Matching coefficients}
\label{sec:app-splitting}
We collect for ease of reference the well-known matching coefficients
which relate
the four and five scheme PDFs. Up to $O(\alpha_s)$
\begin{equation}
  K_{ij}(z,Q^2) = \delta_{ij}\delta(1-z) +
  \alpha_s(Q^2)\,K_{ij}^{(1)}(z,Q^2) +{\cal O}(\alpha_s^2)\,.
\end{equation}
so that
\begin{equation}
  K_{ij}^{-1}(z,Q^2) = \delta_{ij}\delta(1-z) -
  \alpha_s(Q^2)\,K_{ij}^{(1)}(z,Q^2) +{\cal O}(\alpha_s^2)\,.
\end{equation}
The only non-zero contributions at order ${\cal O}(\alpha_s)$
are the heavy quark-heavy quark and the heavy quark-gluon
matching functions, which are respectively given by
\begin{equation}
  \label{eq:ks}
  \begin{split}
    K_{bb}^{(1)}\left(x,\frac{Q^2}{\mu_b^2}\right) & = \frac{C_F}{2\,\pi}{\left\{P_{qq}(x)\left[
          \ln{\frac{Q^2}{\mu_b^2}} -2\ln(1-x)-1 \right] \right\}}_{+} \\
    K_{bg}^{(1)}\left(x,\frac{Q^2}{\mu_b^2}\right) & = \frac{T_R}{2\,\pi}
    P_{qg}(x)\,\ln{\frac{Q^2}{\mu_b^2}}
  \end{split}
\end{equation}
where 
\begin{equation}
  P_{qg}(x) = \left(1\,-\,2\,x\,+\,2\,x^2\right)\quad {\rm and }\quad
  P_{qq}(x) = \frac{2}{1-x} -\left( 1+x \right)\,.
\end{equation}

\section{Massive coefficient functions}
\label{sec:app-coeff}
In this Appendix we summarize  the computation of the
coefficient functions in the massive scheme and of their massless
limit up to $O(\alpha_s)$. The NLO corrections are computed using
the extension of
Catani-Seymour subtraction for massive initial states developed
in Ref.~\cite{Dittmaier:1999mb} and extended to QCD
in Ref.~\cite{Krauss:2017wmx}. This way of preforming the computation 
has the main advantage of following closely that of the
five-flavor massive scheme, so that a direct comparison is much easier
at the analytic level. Indeed, strictly speaking because of
Eq.~(\ref{eq:genfonll}) the massless limit is not needed. However, we
have computed it explicitly in order to check that it matches the
massless-scheme result (thereby verifying Eq.~(\ref{eq:canc})
explicitly), and also in order to produce Fig.~\ref{fig:mub-var},
which provides a further consistency check. Another advantage of this
way of performing the computation (though we do not use it here) is
that it allows for the computation of
the fully differential cross section in this scheme. 

\subsection{Leading order}
The leading order partonic cross section for the production of a Higgs
boson, accounting for the mass of the initial state $b$ and $\bar{b}$,
is given by
\begin{equation}
  \label{eq:bbh-lo}
  \hat{\sigma}_{0}(xs) = \left( \frac{g_{b\bar{b}H}^2\,\beta_0\,\pi}{6}
  \right)\delta(xs -m_H^2) = \sigma_0 \,x\,\delta\left( x-\frac{m_H^2}{s} \right)
\end{equation}
where 
\begin{equation}
  \label{eq:s0}
  \sigma_0 = \frac{g_{b\bar{b}H}^2 \,\beta_0 \, \pi}{6\,m_H^2}\,,
  \quad {\rm and}\quad  \beta_0 = \sqrt{1-\frac{4\,m_b^2}{m_H^2}}\,.
\end{equation}
where $g_{b\bar{b}H}$ is the coupling of the $b$ quark to the Higgs
boson, obtained as the mass of the quark divided by vacuum expectation
value of the Higgs sector:
\begin{equation}
  g_{b\bar{b}H} = \frac{m_b}{v}\,.
\end{equation}
In the following we will also use the notation
\begin{equation}
  {\cal B}(x) \equiv \hat{\sigma}_{0}(xs)\,,\quad {\rm and} \quad
  {\cal B} \equiv \hat{\sigma}_{0}(s)\,.
\end{equation}

\subsection{Next-to-leading order: $b\bar{b}$-channel}
The next to leading order corrections to the Higgs production in
bottom quark fusion consist in virtual corrections (${\cal V}$) to the
left diagram of Fig.~\ref{fig:massive-4fs}, as well as of real
emission corrections (${\cal R}$) , represented by the central diagram of
Fig.~\ref{fig:massive-4fs}.
Both this contributions are separately divergent when the additional
gluon, real or virtual, becomes soft, though the final result remains
finite. In order to handle these soft divergences we employ the
subtraction scheme defined in~\cite{Krauss:2017wmx}. This implies that
we need two more ingredients: a subtraction term, ${\cal S}$, and its
integral over the gluon phase space, ${\cal I} = \int{\rm d}\Phi_g{\cal S}$.
Our final result is then given by:
\begin{equation}
  \hat{\sigma}_{\rm NLO} = \int{\rm d}\Phi_1 {\cal B} +{\cal V}+{\cal
    I}+  \int{\rm d}\Phi_2 {\cal R} -{\cal S}\,.
\end{equation}

\subsubsection{Real corrections, and subtraction term}
The real emission partonic differential cross section, is given by
\begin{equation}
 \int {\rm d} \Phi_{2}  {\cal R} =\int {\rm d} \Phi_{2}  \left| \overline{{\cal M}}_{b\bar{b}Hg} \right|(s,t,u)\,,
\end{equation}
where 
\begin{equation}
  {\rm d} \Phi_{2}=\frac{1}{32\,\pi\,\beta\,s} {\rm d}\cos\theta\,
  \Theta(1+\cos\theta )\,\Theta(1-\cos\theta)\, , \quad{\rm and}\quad
  \beta=\sqrt{1-\frac{4\,m_b^2}{s}}\,,
\end{equation}
and
\begin{equation}
  \label{eq:real}
  \begin{split}
    \left| \overline{{\cal M}}_{b\bar{b}Hg} \right|(s,t,u) = &
    \frac{4}{3}\pi g_{b\bar{b}H}^2C_F\alpha_s
    \left\{
      \left( s-m_H^2\right)\left[ \frac{1}{m_b^2-t}
        + \frac{1}{m_b^2-u}\right]
    \right.\\
    &
    \left.+ (m_H^2-4\,m_b^2)
      \left[
        \frac{2\left( s-2\,m_b^2 \right)}{(m_b^2-t)(m_b^2-u)} -
        \frac{2\,m_b^2}{(m_b^2-t)^2} -
        \frac{2\,m_b^2}{(m_b^2-u)^2}
      \right]
    \right\}\,.
  \end{split}
\end{equation}
The Mandelstam variables in terms of scalar
products and $\cos\theta$ are given by
\begin{equation}
  \left\{
    \begin{split}
      t &= m_b^2 - \frac{s-m_H^2}{2}\left( 1-\beta\,\cos\theta\right)\\
      u &= m_b^2 - \frac{s-m_H^2}{2}\left( 1+\beta\,\cos\theta\right) 
    \end{split}
  \right.\,.
\end{equation}

In order to remove the soft divergence which appears in the  $s\rightarrow
m_H^2$ limit we need to construct a suitable  subtraction term. Using
the relevant 
equations in Ref.~\cite{Krauss:2017wmx} we find
\begin{equation}
  \label{eq:sub}
  {\cal S} =
  \frac{2}{3}\,\pi \,\alpha_s\,C_F\,g_{b\bar{b}H}^2\,\beta_0^2\,m_H^2
  \frac{1}{\tilde{x}}
  \left[\frac{2}{m_b^2-t}\left( P_{qq}(\tilde{x})-\frac{2\,\tilde{x}\,m_b^2}{m_b^2-t} \right)
    +\frac{2}{m_b^2-u}\left( P_{qq}(\tilde{x})-\frac{2\,\tilde{x}\,m_b^2}{m_b^2-u} \right)\right]
\end{equation}
where
\begin{equation}
  \tilde{x} = \frac{m_H^2-2\,m_b^2}{s-2\,m_b^2}\,.
\end{equation}

Combining Eqs.~(\ref{eq:real}) and~(\ref{eq:sub}) and factoring
the trivial $\frac{\alpha_s\,C_F\,\sigma_0}{\pi}$ dependence we get
\begin{equation}
  \begin{split}
    \frac{\alpha_s\,C_F\,\sigma_0}{\pi}
    \int{\rm d}\Phi_2 \left[{\cal R} -{\cal S}\right] & =
    \frac{\alpha_s\,C_F\,\sigma_0}{\pi}\frac{m_b^2}{2}
    \int_{-1}^1{\rm d}\cos\theta
    \left[\frac{s\,(s-m_H^2)^2}{(m_H^2-2\,m_b^2)(m_b^2-t)(m_b^2-u)}
    \right]\\
    & = -\frac{\alpha_s\,C_F\,\sigma_0}{\pi}
    \frac{1}{\beta_0}\left(\frac{ 1-\beta^2 }{\beta^2}\right)
    \frac{x}{\left( 1 - 2\,x - \beta^2\right)}\ln d\,,
  \end{split}
\end{equation}
where we defined
\begin{equation}
  d \equiv \frac{1+\beta}{1-\beta}\,,
  \quad {\rm and}\quad x\equiv \frac{m_H^2}{s}\,.
\end{equation}

\subsubsection{Virtual corrections, and integrated subtraction term}
QCD virtual corrections to the Born process in this simple case
completely factorize in a vertex form factor:
\begin{equation}
  {\cal V} = \frac{\alpha_s\,C_F}{\pi}{\cal B}\,\delta_g\,,
\end{equation}
with
\begin{equation}
  \delta_g = -1 - L_{\lambda} +
  \frac{(1-\beta_0^2)}{\beta_0} \ln d_0 
  - \frac{1+\beta_0^2}{2\,\beta_0}\left[-\ln d_0\,L_{\lambda} + \ln^2d_0+{\rm Li}_2\left(
      1-\frac{1}{d_0} \right) -\frac{\pi^2}{2} \right]\,,
\end{equation}
where
\begin{equation}
  L_{\lambda} \equiv \frac{1}{\epsilon} + \ln\frac{4\,\pi\,\mu_R^2}{m_b^2}
  +{\cal O}(\epsilon^2)\,.
\end{equation}

The integrated subtraction term ${\cal I}$ is obtained by
integrating  ${\cal S}$, Eq.~(\ref{eq:sub}),  over the phase space of
the emitted gluon. This term can be separated into two pieces: a term
proportional to 
$\delta(1-x)$, which contains the singularity,
and a plus distribution:
\begin{equation}
  {\cal I} = \delta(1-x)\,I + \left\{ {\cal G}(x) \right\}_+\,,
\end{equation}
where
\begin{equation}
  \begin{split}
    I  = &2 +L_{\lambda}- \ln{\frac{(1+\beta_0^2)^2}{1-\beta_0^2}} +
    \frac{1-3\,\beta_0^2}{4\beta_0}\ln d_0\\
     &+\frac{1+\beta_0^2}{2\,\beta_0}
    \left[
      \frac{1}{2}\ln^2 d_0
      -\ln d_0\,\ln {\frac{4\,\beta_0^2}{(1+\beta)^2}}
      - L_{\lambda}\,\ln d_0 -1
      +2\,{\rm Li}_2\left(\frac{1}{d_0} \right)-\frac{\pi^2}{3}
    \right]\,,
  \end{split}
\end{equation}
and
\begin{equation}
  \{{\cal G}(x)\}_+ =  {\left\{P_{qq}(x)\left[ \frac{1+\beta^2}{2\beta}
        \ln{d} -1 \right] +
      \left( 1-x \right)\right\}}_{+}\,.
\end{equation}
\subsubsection{Final formulae, mass and PDF renormalization}
We now combine the various partial results obtained in the previous
subsections  into the  full expression for the $b\bar{b}$-channel
coefficient functions. First, however, we need to adjust 
$b$-quark mass and the PDFs.
Renormalization of the  $b$ mass leads to the   replacement
\begin{equation}
  g_{hb\bar{b}}^2 = g_{hb\bar{b}}^2(\mu_R^2) \left(
    1-\frac{\alpha_s\,C_F}{\pi}\left(
        \frac{3}{2}\ln\frac{m_b^2}{\mu_R^2}-2  \right) \right)\,.
\end{equation}
in $\sigma_0$, Eq.~\eqref{eq:s0}.

The massive $b$ PDF is free of collinear singularities and thus it
does
not have to
undergo subtraction: indeed it is scale independent.
However, we must perform the change of
renormalization scheme Eq.~\eqref{eq:change-of-scheme} which relates
the massive and massless schemes.
Up to ${\cal O}(\alpha_s)$ we get
\begin{equation}
  B_{b\bar{b}}\left(x,\mu_R^2,\mu_F^2 ,\mu_b^2\right) =
  \,\left[ \sigma_0(\mu_R^2) \delta(1-x) + \alpha_s(\mu_R^2)
    B_{b\bar{b}}^{(1)}\left(x,\mu_R^2,\mu_F^2 ,\mu_b^2\right) \right]
  +{\cal O}(\alpha_s^2)
\end{equation}
where
\begin{equation}
  \begin{split}
  B_{b\bar{b}}^{(1)}\left(x,\mu_R^2,\mu_F^2 ,\mu_b^2\right) =
  \frac{\sigma_0(\mu_R^2)\,C_F}{\pi}&
  \left\{
    \left[
      \frac{3}{2}\ln\frac{\mu_R^2}{\mu_b^2}+2 + I + \delta_g
    \right]\delta(1-x) \right.\\
  &\left.
   + \int_0^1{\rm d}z \{{\cal G}(z) - 2\,K_{b\bar{b}}^{(1)}(z)\}_+z\,\delta(z-x)
   +  \int{\rm d}\Phi_2 \left[{\cal R} -{\cal S}\right] \right\}\,.
  \end{split}
\end{equation}

Performing the $z$ integration gives the final result
\begin{equation}
  \label{eq:b1-massive}
  \begin{split}
    B_{b\bar{b}}^{(1)}&\left(x,\mu_R^2,\mu_F^2 ,\mu_b^2\right) =
    \frac{\sigma_0(\mu_R^2)\,C_F}{\pi}
    \Biggl\{
    \delta(1-x)
    \left[
      \xi -2 + \frac{3}{2}\left(\gamma_0\ln
        \frac{(1+\beta)^2}{4} -\gamma_0\ln \frac{m_H^2}{m_b^2}+
        \ln\frac{\mu_R^2}{\mu_F^2} \right)
    \right]\\
  +&\,4\,{\cal D}_1(1-x) 
  +  2
  \left[
    \gamma\ln\frac{(1+\beta)^2}{4}+\gamma\ln\frac{m_H^2}{m_b^2}+
    \ln{\frac{\mu_b^2}{\mu_F^2}}
  \right]\,
  {\cal D}_0(1-x)\\
  -& (2 + x + x^2)\left[\gamma\ln\frac{(1+\beta)^2}{4}
    +\gamma\ln\frac{m_H^2}{m_b^2}-\gamma\ln x+
    \ln{\frac{\mu_b^2}{\mu_F^2}}
    +2\ln(1-x)\right]+ x\,(1-x) \\
  -&  \frac{2\,\gamma\,\ln x}{1-x} -
    \frac{1}{\beta_0}\left(\frac{ 1-\beta^2}{\beta^2}\right)
    \frac{x}{\left( 1 - 2\,x - \beta^2\right)}\ln d \Biggr\}\,,
  \end{split} 
\end{equation}
where
\begin{equation}
  \xi = 1+\ln\left(\frac{1-\beta_0^2}{(1+\beta_0)^2}\right)+
  \frac{\left(5-7\beta_0^2\right)}{4\,\beta_0}\ln d_0+
  \frac{\left(\beta_0^2+1\right)}{\beta_0} \left( 2\,{\rm
      Li}_2\left(\frac{1}{d_0}\right)+\frac{\pi ^2}{6} -\ln d_0
    \ln\frac{4\beta_0^2} {(1+\beta_0^2)(1+\beta_0)}\right)\,,
\end{equation}
and
\begin{equation}
  \gamma = \frac{1+\beta^2}{2\,\beta}\,,\quad \gamma_0 =
  \frac{1+\beta_0^2}{\,2\beta_0}\, \quad \text{ and }
  \quad {\cal D}_n(x) ={\left( \frac{\ln^n(1-x)}{1-x} \right) }_{+}\,.
\end{equation}

\subsubsection{Massless limit}
The massless limit of the $b\bar{b}$-channel can be computed directly
from Eq.~\eqref{eq:b1-massive},
by setting $\beta=1$ everywhere except in the logarithms, where
one can use the simple expansion
\begin{equation}
  \label{eq:beta}
  \beta \sim 1\,-\,\frac{2\,x\,m_b^2}{m_H^2} + {\cal O}\left( \frac{m_b^4}{m_H^4} \right)\,.
\end{equation}
We get
\begin{equation}
  \label{eq:b1-massless}
  \begin{split}
    B_{b\bar{b}}^{(1),(0)}&\left(x,\mu_R^2,\mu_F^2,\mu_b^2 \right) = 
    \frac{\alpha_s \, C_F\,\sigma_0(\mu_R^2)}{\pi}
    \Biggl\{
      \delta(1-x)\left[ -1+ \frac{\pi^2}{3} + \frac{3}{2}
          \ln\frac{\mu_R^2}{\mu_F^2} \right] +\,4\,{\cal
        D}_1(1-x) \\
      +&  2\left( 
      \ln{\frac{m_H^2}{\mu_F^2}}+ \ln{\frac{\mu_b^2}{m_b^2}}\right)\,{\cal D}_0(1-x)
      -\frac{2\ln x}{1-x} - (2 + x + x^2)\left[
        \ln\frac{m_H^2}{\mu_F^2} +\ln{\frac{\mu_b^2}{m_b^2}} + \ln \frac{(1-x)^2}{x}\right]+ x\,(1-x)  \Biggr\}\,.
  \end{split} 
\end{equation}
As it can be easily verified, this exactly corresponds to its massless
scheme equivalent, which can be found in Eq.~(A6) of Ref.~\cite{Harlander:2003ai}.
\subsection{Next-to-leading order: $bg$-channel}
In the presence of initial-state massive quarks, the cross-section for
the $bg$-channel is free of  soft or 
collinear divergences, and no subtraction is accordingly
necessary. Also in this case, however, we must  perform the scheme
change Eq.~\eqref{eq:change-of-scheme}. We get
\begin{equation}
  \begin{split}
    B_{bg}^{(1)}(x,\mu_R^2,\mu_F^2,\mu_b^2) & = \hat{\sigma}_{bg}(x,\mu_R^2)
    - \alpha_s\,\int_0^1{\rm d}z\, K_{bg}^{(1)}(z,\mu_F^2)\sigma(zs)\,\\
    & = \left. \hat{\sigma}_{bg}(x,\mu_R^2) - \frac{\alpha_s\,
        T_R\,\sigma_0}{\pi}\left[ \frac{x}{2}\,P_{qg}(x)
        \ln{\frac{\mu_F^2}{\mu_b^2}}\right]\,\right|_{x=\tfrac{m_H^2}{s}},
  \end{split}
\end{equation}
where
\begin{equation}
  \hat{\sigma}_{bg}(x,\mu_R^2) = \int{\rm d}\Phi_2^{(b)}
  \left| \overline{{\cal M}}_{bgHb} \right|^2(s,t,u)\,,
\end{equation}
and the subscript $(b)$ in $\Phi_2^{(b)}$ denotes the fact that now
the phase-space has a massive $b$ instead of a massless gluon, in the
final state. The color- and helicity-averaged square matrix element,
can be obtained from Eq.~\eqref{eq:real} using crossing symmetry. In
addition, we have to take into account that the gluon can have 8
possible colors (as opposed to 3 for a quark),
\begin{equation}
  \left| \overline{{\cal M}}_{bgHb} \right|^2(s,t,u)=
  - \frac{3}{8} \left| \overline{{\cal M}}_{b\bar{b}Hg} \right|^2(t,s,u)\,,
\end{equation}
where now the Mandelstam invariants are given by
\begin{equation}
  \left\{
    \begin{split}
      t &= 2\,m_b^2 +
      \frac{s}{32}\left((5-\beta^2)(\beta^2+4\,x-5) -
        (3+\beta^2)\,\Lambda\,\cos\theta\right)\\
      u &= m_b^2 +\frac{s}{32}\left((5-\beta^2)(\beta^2+4\,x-5) +
        (3+\beta^2)\,\Lambda\,\cos\theta\right), 
    \end{split}
  \right.\,
\end{equation}
where
\begin{equation}
  \Lambda = \sqrt{\left( 3+\beta^2 \right)^2 + 16 x^2 -
    8\,x\,\left(5-\beta^2\right)}\,,
\end{equation}
while the phase-space ${\rm d}\Phi_2^{(b)}$ is given by
\begin{equation}
  {\rm d}\Phi_2^{(b)} = \frac{\Lambda\,x}{32\,\pi\,\left( 3+\beta^2
    \right)\,m_H^2}\,{\rm d}\cos\theta\,
  \Theta(1+\cos\theta )\,\Theta(1-\cos\theta)\,.
\end{equation}

Performing the $\cos\theta$ integration gives
\begin{equation}
  \begin{split}
    \hat{\sigma}_{bg}(x,\mu_R^2) =&
    \frac{\alpha_s\,T_R\,\sigma_0(\mu_R^2)}{\pi}\frac{  x}{16 \beta_0
      \left(\beta^2+3\right)^3}\\
    &\times\Biggl\{- 64 \left(9 \beta^4+(40
     x-42) \beta^2 +8 x (4 x-9)+49\right){\rm arctanh}\left(\frac{\Lambda }{\beta^2+4
       x-5}\right)   \\
    &\frac{4096\,\Lambda\, \left(1-\beta^2 \right)
      \left(\beta^2+x-1\right)} {\left(-\Lambda +\beta^2+4 x-5\right)
      \left(\Lambda +\beta^2+4 x-5\right)} +
    \Lambda\,\left(5-\beta^2\right) \left(\beta^4+(4 x+22) \beta^2+44 x-71\right)
  \Biggr\}\,.
  \end{split}
\end{equation}

\subsubsection{Massless limit}
As in the case of the $b\bar{b}$ channel, taking the massless limit
requires setting $\beta=1$ everywhere except in the logarithms where
one can use Eq.~\eqref{eq:beta}, which gives
\begin{equation}
  \label{eq:mlimbg}
  B_{bg}^{(1),(0)}(x,\mu_R^2,\mu_F^2,\mu_b^2) =
  \frac{T_R}{\pi}\left\{  
    \frac{x}{2}\,P_{qg}(x)\left[
      \ln\left(\frac{(1-x)^2}{x}\right)+\ln\frac{m_H^2}{\mu_F^2}+
      \ln{\frac{\mu_b^2}{m_b^2}} \right]-\frac{x}{4}(1-x)(3-7\,x) \right\}\,.
\end{equation}
Once again, one can explicitly check that this exactly corresponds to
its massless scheme counterpart, which can be found in Eq.~(A9) of Ref.~\cite{Harlander:2003ai}.
\renewcommand{\em}{}
\bibliographystyle{UTPstyle}
\bibliography{bbh-intrinsic}

\end{document}